\documentclass[a4paper,twocolumn,11pt]{quantumarticle}
\pdfoutput=1
\usepackage[utf8]{inputenc}
\usepackage[english]{babel}
\usepackage[T1]{fontenc}
\usepackage{amsmath}
\usepackage{hyperref}
\usepackage{epsfig}
\usepackage{amssymb}
\usepackage{amsmath}
\usepackage{amsfonts}
\usepackage{mathtools}
\usepackage{bm}
\usepackage{tikz}
\usetikzlibrary{quantikz}
\usepackage{physics}
\usepackage{geometry}
\usepackage{enumitem}
\usepackage{dsfont}
\usepackage{graphicx,grffile}

\usepackage[ruled,vlined]{algorithm2e}
\usepackage[toc,page]{appendix}
\usepackage[caption=false]{subfig}

\hypersetup{colorlinks=true, linkcolor=blue, citecolor=blue, urlcolor=blue, unicode=true}
\usepackage{cleveref}
\Crefname{equation}{Eq.}{Eqs.}

\renewcommand{\grad}{\text{grad }}
\newcommand{\M}{\mathcal{M}}
\newcommand{\N}{\mathcal{N}}
\newcommand{\R}{\mathbb{R}}

\newcommand{\SU[1]}{\text{SU}(#1)}
\newcommand{\su[1]}{\mathfrak{su}(#1)}

\newcommand{\suloc[1]}{\mathfrak{su}_{\text{loc}_#1}}
\newcounter{circuit}[section]

\Crefname{table}{table}{tables}
\Crefname{table}{Table}{Tables}
\Crefname{circuit}{circuit}{circuits}
\Crefname{circuit}{Circuit}{Circuits}

\begin{document}

\title{Optimizing quantum circuits with Riemannian gradient flow}

\author{Roeland Wiersema}
\affiliation{Xanadu, Toronto, ON, M5G 2C8, Canada}
\affiliation{Vector Institute, MaRS  Centre,  Toronto,  Ontario,  M5G  1M1,  Canada}
\affiliation{Department of Physics and Astronomy, University of Waterloo, Ontario, N2L 3G1, Canada}
\author{Nathan Killoran}
\affiliation{Xanadu, Toronto, ON, M5G 2C8, Canada}
\maketitle

\begin{abstract}
    Variational quantum algorithms are a promising class of algorithms that can be performed on currently available quantum computers. In most settings, the free parameters of a variational circuit are optimized using a classical optimizer that updates parameters in Euclidean geometry. Since quantum circuits are elements of the special unitary group, we can consider an alternative optimization perspective that depends on the structure of this group. In this work, we investigate a Riemannian optimization scheme over the special unitary group and we discuss its implementation on a quantum computer. We illustrate that the resulting Riemannian gradient-flow algorithm has favorable optimization properties for deep circuits and that an approximate version of this algorithm can be performed on near-term hardware.
\end{abstract}

\section{Introduction}

With quantum computing hardware still in its infancy, variational quantum algorithms offer a way to probe the power of noisy intermediate-scale quantum (NISQ) devices \cite{Peruzzo2014vqe, Farhi2014qaoa}. In a typical setup, one calculates gradients with respect to gate parameters in a quantum circuit to minimize a cost function that depends on the variational state. Since these approaches often involve minimizing non-convex cost functions, the choice of optimizer can greatly affect the result \cite{Sung2020compareopt}. Unlike in deep learning, where backpropagation can remain effective despite a large number of parameters, calculating gradients in a variational quantum circuit quickly becomes inefficient. This is due to the fact that the gradients for single parameters cannot be calculated concurrently, but require additional circuit evaluations for each parameter \cite{Mitarai2018grads,Schuld2019grads}.

Gradient-based methods can be improved by considering additional structure of the model under consideration. For instance, when dealing with a statistical model, one can make use of the Fisher information to quantify the statistical distance between probability distributions \cite{Fisher1922info}. This induces a metric on parameter space, which provides the direction of steepest descent with respect to the information geometry \cite{Amari1998natgrad}. The resulting gradient is called the natural gradient. Optimization schemes that make use of the natural gradient are used in deep learning \cite{Pascanu2014natgrad, Zhang2018l} and variational quantum Monte Carlo \cite{Sorella1998stochreconf}, but can also be extended to variational quantum circuits with the quantum natural gradient, where the distance between rays in Hilbert space provides a analogue of the Fisher information \cite{Stokes2020quantumnatural, Katabarwara2021rieman}. 

Optimization algorithms that rely on the Fisher information fall into the category of Riemannian optimization algorithms \cite{Becigneul2018riemannian, Udriste1994riemopt}. However, they are limited to optimizing over a real parameter space $\R^n$ with a non-Euclidean metric. Riemannian optimization has a much broader application: we can consider minimizing a function over a differentiable manifold $\M$ equipped with a non-degenerate, positive metric. This construction is more general, and allows one to take the structure of the manifold into account during the optimization. Such applications have been considered in the context of quantum control \cite{SchulteHerbruggen2010gradflow, Helmke1994optdyn, LuchnikovRGautodif,HelmkeAug2002riemanngrad,Glaser1998nmrrieman,Huang2018Grassmann}, or optimization of neural networks  \cite{Fiori2005geodesiclearning, Fiori2010natgradpseudRG, Wisdom2016unitarynns}. In the quantum circuit setting, the Riemannian manifold perspective has been considered to study the computational complexity of constructing specific circuits by approximating geodesics on the unitary group  \cite{Nielsen2006rieman, Dowling2008geom}. 

In this work, we introduce the optimization of quantum circuits over the special unitary group $\SU[p]$ using Riemannian gradient flows \cite{SchulteHerbruggen2010gradflow}. We show the resulting algorithm can produce quantum circuits with favorable optimization properties but which may be exponentially deep. To obtain a practically feasible circuit optimizer, we make approximations that keep gate costs under control. We explore several toy problems to illustrate the properties of the resulting exact and approximate Riemannian gradient flow.

In \Cref{sec:background}, we introduce the necessary theory of gradient flows on the special unitary group. Then in \Cref{sec:exact}, we show how these flows can be adapted to the variational quantum circuit setting. To make these algorithms practical, we have to consider approximations in \Cref{sec:approx}, for which we present numerical results for some toy models. In addition, we argue how some of the literature on adaptive variational approaches can be re-contextualized from the Riemannian point of view. In \Cref{sec:conclusion} we summarize our results.

\section{Background \label{sec:background}}

\subsection{Gradient flows in quantum circuits}

An archetypal example of a widely used gradient flow in quantum computing is the Variational Quantum Eigensolver (VQE) \cite{Peruzzo2014vqe}. Consider the cost function $\mathcal{L}: \R^n \to \R$,
\begin{align}
    \mathcal{L}(\theta) = \Tr{H U(\theta) \rho_0 U(\theta)^\dag} \equiv \expval{H}_\theta \label{eq:vqe},
\end{align}
where $U(\theta)$ is a parameterized quantum circuit, $H$ a Hamiltonian whose ground state we want to approximate and $\theta \in (0,2\pi)^n$ is a vector of gate parameters. Here, $\rho_0=\ketbra{\psi_0}{\psi_0}$ is some initial state of the system, usually taken to be the zero state $\ketbra{0}$. We are interested in minimizing $\mathcal{L}(\theta)$ with respect to the parameters $\theta$. 

To solve the optimization problem $\min_\theta \mathcal{L}(\theta)$, we can consider the flow
\begin{align}
    \dot{\theta} = \nabla_\theta \mathcal{L}(\theta), \label{eq:gradflow}
\end{align}
where $\nabla_\theta = \sum_i^n \partial_{\theta_{i}}$ is the standard gradient operator. \Cref{eq:gradflow} provides a differential equation for the evolution of the parameters based on the gradient of the function at a point $\theta$. This flow equation can be discretized as
\begin{align}
    \theta_{k+1} = \theta_k - \epsilon \nabla_\theta\mathcal{L}(\theta),\label{eq:gradflow_vqe}
\end{align}
where $\epsilon$ is the step size that controls the precision of the discretization.
Using this equation to update the parameters of $\mathcal{L}(\theta)$ is called steepest descent, since we follow the gradient of the function to a minimum.

To understand why this works, we can look at the level curves of $\mathcal{L}$, i.e., curves through parameter space where the function $\mathcal{L}$ is constant. We can define such a curve as $\gamma:(-a ,a)\to \R^n$ with $\gamma(0) = \theta$ such that $\mathcal{L}(\gamma(t)) = \text{constant}$. Differentiating with respect to $t$ then gives
\begin{align}
    \sum_i \partial_{\theta_{i}}\mathcal{L}(\gamma(t))\gamma_i'(t)\bigg|_{t=0} = 0,
\end{align}
which we identify as $\nabla \mathcal{L}\cdot v$, the gradient of $\mathcal{L}$ in the direction of $v=\gamma'(t)$. In other words, the gradient of a function produces a vector orthogonal to the level curves through a point. As a result, infinitesimal steps in the direction of the gradient will decrease the function's value until we reach a local minimum \cite{Lee2016conv}. 

One issue with VQE is that the parameterization of the variational circuit $U(\theta)$ is an arbitrary choice that we have to make. This implies that one must try different ans\"atze and assume that the state of interest can be expressed with the chosen ansatz. Moreover, due to the non-convexity of the cost landscape, we have no guarantees that our optimizer can find a good approximation to the desired state.



\subsection{Gradient flows on Lie groups}

\begin{figure*}
    \subfloat[\label{fig:append}]{
    \centering
    \includegraphics[width=0.45\textwidth]{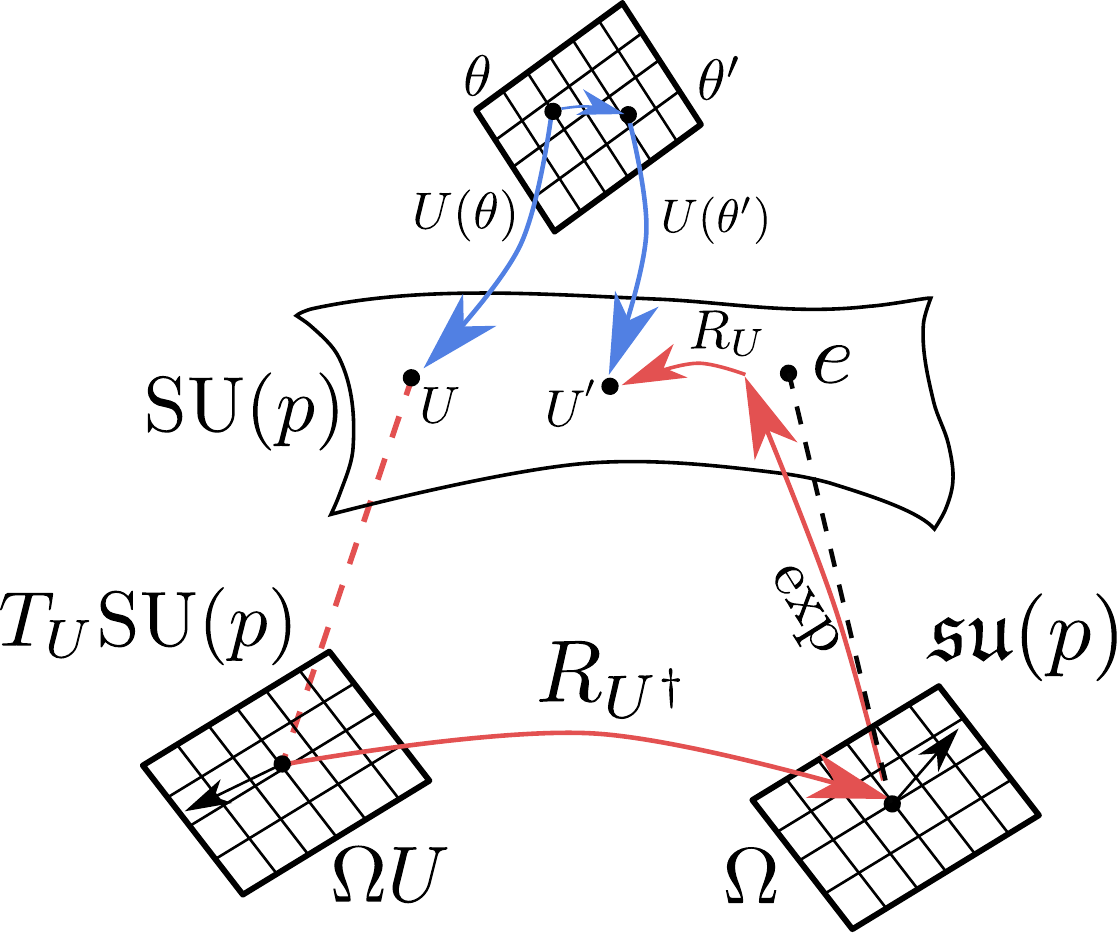}
    }
    \hspace{5mm}
    \subfloat[\label{fig:opt_diff}]{
    \centering
    \includegraphics[width=0.45\textwidth]{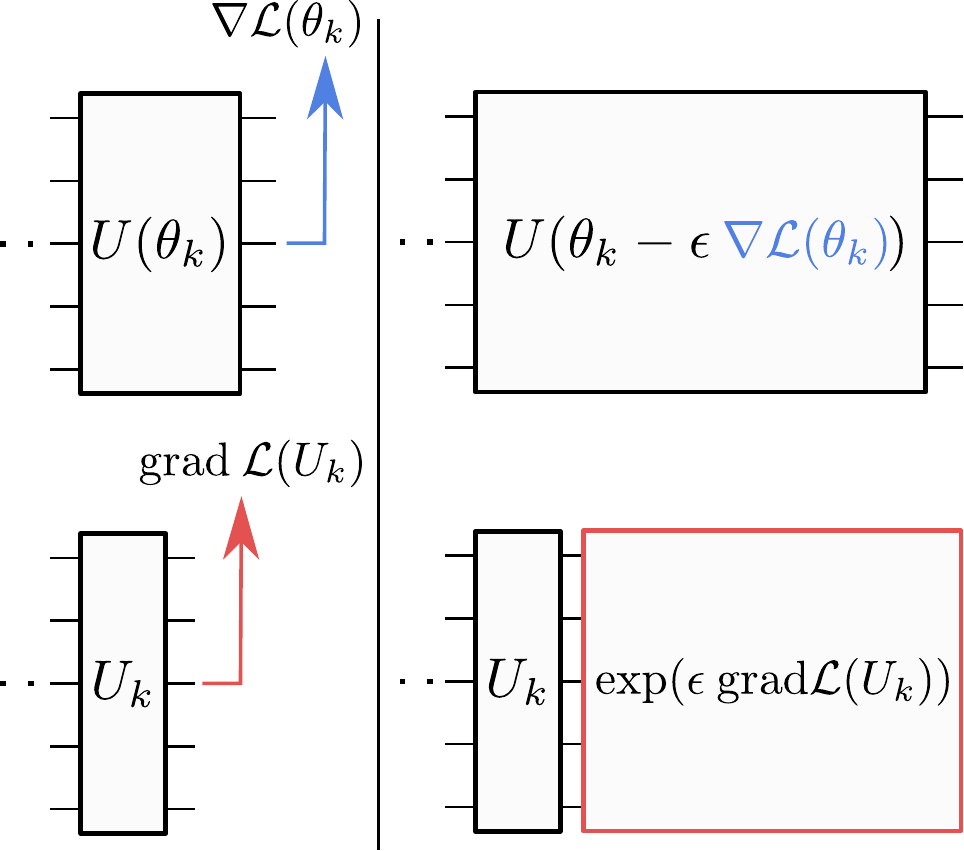}
    }
    \label{fig:euc_vs_lie}
    \caption{Difference between the Riemannian gradient flow and Euclidean gradient flow. (a) In blue, we have a mapping from real parameters $\theta\in\R^n$ to a unitary in $U\in\SU[p]$. Gradient updates in the parameter space from $\theta$ to $\theta'$ result in a new unitary $U'$ on the group. In red, we first obtain the Riemannian gradient at $U$ in the tangent space $T_U \SU[p]$. Since the Riemannian gradient can be written as $\Omega U$ with $\Omega\in \su[p]$, we can move to the Lie algebra $\su[p]$ by multiplying the Riemannian gradient with $U^{\dagger}$ from the right. Then, the exponential map and subsequent right multiplication with $U$ projects the Riemannian gradient back onto the manifold which results in a new unitary $U'$. (b) At the top, we see how a standard gradient flow optimizes a quantum circuit: The circuit stays fixed and the gradient is calculated via the parameter-shift rule. Next, the free parameters describing the unitary are updated via gradient descent. In the bottom figure, we see that a step of the Riemannian optimization corresponds to appending a new unitary to the original circuit. }
\end{figure*}

A quantum circuit $U$ is a unitary operation that is an element of the special unitary group $\SU[p]$. What if instead of considering the optimization problem over $\R^n$ for a particular parameterization $U(\theta)$, we directly optimize over $\SU[p]$?

For such a construction to make sense, we need to introduce a gradient on $\SU[p]$. Since $\SU[p]$ is a finite-dimensional Lie group, it carries a differentiable manifold structure. We can therefore use differential geometry to define a gradient on the group. In particular, a $p$-dimensional manifold $\mathcal{M}$ is a set that locally looks like $\R^p$. This local description is given by charts, which smoothly map open subsets of the manifold onto coordinate patches in $\R^p$. If all charts between two subsets of the manifold are compatible, the manifold is differentiable (see Appendix \ref{app:manifolds}).

The tangent space $T_U \SU[p]$ of the manifold at a point $U$ is a vector space that consists of a collection of vectors $\Omega\in T_U \SU[p]$ that provide the possible directions one can move in on the manifold from point $U$. The tangent vectors $\Omega$ can be defined as derivatives of curves going through the point $U$ (see Appendix \ref{app:tangent_space}). For example, on a sphere, the tangent space at a point $p$ consists of a plane tangent to $p$ .

The introduction of an inner product on the tangent space turns the manifold into a Riemannian manifold, with well-defined notions of angles and distance (see Appendix \ref{app:riemannian_manifolds}). Given this metric, the Riemannian gradient $\grad\:\mathcal{L}(U)$ of a function $\mathcal{L}:\SU[p]\to\R$ at $U$ can be constructed by satisfying two conditions:

1) The Riemannian gradient $\grad\:\mathcal{L}(U)$ at a point $U$ must be an element of the tangent space $T_U\SU[p]$. This ensures that it is always tangential to the manifold at each point, hence $\grad\:\mathcal{L}(U) \in T_U\SU[p]$.

\begin{figure}[htb!]
    \centering
    \includegraphics[width=\columnwidth]{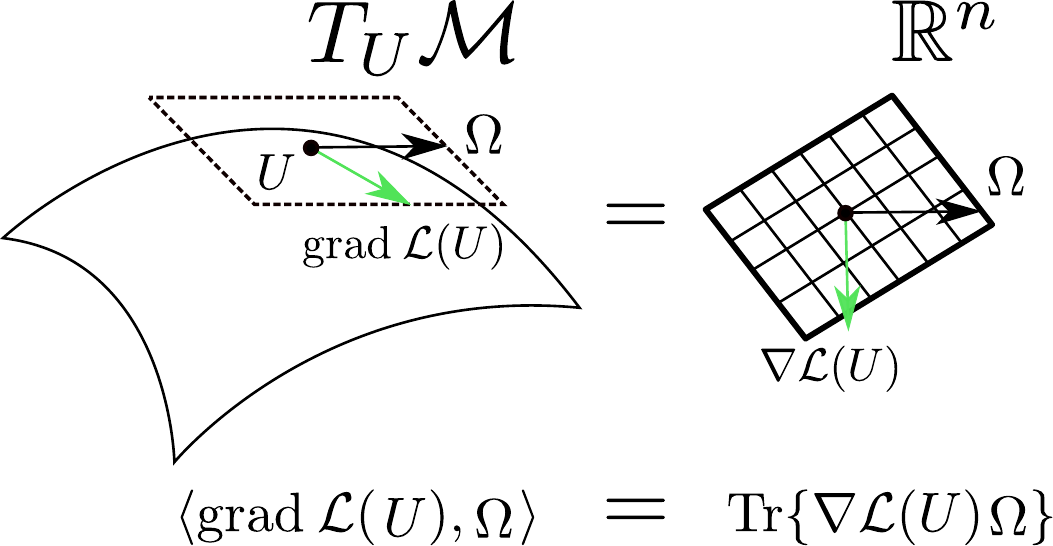}
    \caption{The compatibility condition. By taking the Euclidean inner product as the reference inner product, we can enforce the invariance of the inner product under a change of metric and explicitly construct $\grad  \mathcal{L}(X)$.}
    \label{fig:tangency}
\end{figure}

2) Because there are many different ways to set coordinates on a manifold, and because the function itself should be invariant under a change of coordinates (i.e., its level curves are at the same locations on the manifold), we need to enforce a coordinate-invariant notion of a gradient. This can be achieved with the compatibility condition,
\begin{align}
    \expval{\grad \:\mathcal{L}(U), \Omega} = \Tr{\nabla \mathcal{L}(U) \Omega}
\end{align}
which expresses the fact that the inner product (under some chosen metric) of the Riemannian gradient with any other tangent vector $\Omega$ is independent of the choice of metric (see \Cref{fig:tangency}) \cite{Fiori2005geodesiclearning, LuchnikovRGautodif}. Since the choice of metric should not matter, we can choose any convenient metric as a reference. The reference inner product on the right hand side is taken to be the Euclidean inner product in the local coordinates $\R^p$.

With the two conditions for the Riemannian gradient given above, we can explicitly construct $\text{grad}\: \mathcal{L}(U)$. 
First, we rewrite \Cref{eq:vqe} as a scalar function on the special unitary group, $\mathcal{L}: \SU[p] \to \R$, to obtain
\begin{align}
    \mathcal{L}(U) = \Tr{H U \rho_0 U^\dag} \label{eq:vqe_u},
\end{align}
where $U\in \SU[p]$. To solve the optimization problem $\min_U \mathcal{L}(U)$, we can can consider the Riemannian gradient flow
\begin{align}
    \dot{U} = \text{grad}\: \mathcal{L}(U).
\end{align}

Next, we realize that the tangent space of $\SU[p]$ at the identity element $X_0=\mathbb{I}$ is given by the Lie algebra $\su[p]$, the set of $p\times p$ skew-Hermitian matrices $\Omega$ with $\Tr{\Omega}=0$. The elements of $T_U\SU[p]$ can then be found by right multiplying an element of the Lie algebra with $U$ (see Appendix \ref{app:su_p}):
\begin{align}
    T_U \SU[p] \coloneqq \{\Omega U| \Omega \in \su[p]\}.
\end{align}
With the notion of a tangent vector on $\SU[p]$, we can enforce the compatibility condition (see Appendix \ref{app:lie_grad_flow}) and derive the resulting Riemannian gradient flow on $\SU[p]$:
\begin{align}
    \dot{U} = \text{grad}\: \mathcal{L}(U) = \comm{U\rho_0 U^\dag}{H} U. \label{eq:riemannian_gradient_flow}
\end{align}
Analogous to the gradient in $\R^n$, the Riemannian gradient flow of \Cref{eq:riemannian_gradient_flow} converges to a critical point of $\mathcal{L}(U)$ on $\SU[p]$ by descending along the level curves of the function \cite{Helmke1994optdyn}. 

Because the commutator $\comm{U\rho_0 U^\dag}{H}$ is itself a skew-Hermitian matrix in the tangent space of $\SU[p]$ at $U$, left multiplication of $U$ with the commutator will in general not keep us on the manifold. We therefore have to use a retraction to project the Riemannian gradient from the tangent space back onto $\SU[p]$ \cite{Fiori2008retractions}. In contrast, for the Euclidean case of \Cref{eq:gradflow} where $\M =\R^n$ this problem does not appear, because the tangent space of $\R^n$ coincides with the manifold at all points: $T_\theta \R^n \cong \R^n$. 

The canonical retraction for our setting is the Lie exponential map $\exp_U:T_U\SU[p] \to \SU[p]$, $\Omega\mapsto \exp_U\{\Omega\}$, so that $\exp_U\{t \Omega\}$ for $t\in [0,1]$ describes a unique geodesic at $U$ with initial {``velocity''} $\Omega \in T_U \SU[p]$. 
The operator $\exp_U$ can be decomposed as follows. We realize that $\grad \:\mathcal{L}(U) = \Omega U$ with $\Omega = \comm{U\rho_0 U^\dag}{H}$, hence right multiplication with the inverse $U^{\dagger}$ yields an element of the Lie algebra. Taking $\exp{\Omega}$ and multiplying with $U$ from the right then produces the retracted gradient, see \Cref{fig:append}. If we discretize \Cref{eq:riemannian_gradient_flow} and perform the retraction, we finally obtain
\begin{align}
    U_{k+1} = \exp{\epsilon \comm{U_k \rho_0 U_k^\dag}{H}} U_k, \label{eq:discr_riemannian_gradient_flow}
\end{align}
where $\epsilon$ is the step size and $U_k\in SU(p)$ the unitary at step $k$.  

If $U_k$ is implemented by a quantum circuit, then left multiplication of $U_k$ with the retracted Riemannian gradient is nothing more than appending a set of gates to that circuit, as illustrated in \Cref{fig:opt_diff}.


To analyze the convergence properties of \Cref{eq:discr_riemannian_gradient_flow}, we rely on the fact that the map $\rho_0\mapsto U_k \rho_0 U_k^\dag$ can be understood as a so-called double bracket flow on the adjoint orbits of the group
\cite{Arvanitoyeorgos2003AnIT, Tam2004doublebracket, Chu1990projgrad}.
Double bracket flows can be used to solve a variety of tasks such as sorting lists \cite{Brockett1991listsort}, describing Toda flows \cite{Bloch1992dbtoda}, or diagonalizing Hamiltonians in many-body physics \cite{Wegner1994hamflow, Kehrein2006dbparticle, Glazek1993renormhams}.  Additionally, they have been studied in the context of quantum gate design \cite{Dawson2008dbflows}. The properties of this optimization scheme are well understood, in particular, if $H$ is non-degenerate there exist exactly $p!$ minima on $\SU[p]$, and $(p-1)!$ global minima. Amazingly, only the global minima are stable attractors of the optimization dynamics, and one can show that almost all points will converge to these minima given a suitable step size \cite{Helmke1994optdyn}. Hence the Riemannian gradient flow is guaranteed to find the ground state of a non-degenerate Hamiltonian $H$.

\section{Exact Riemannian gradient flow in quantum circuits \label{sec:exact}}

It should come as no surprise that an implementation of the Riemannian gradient flow on a quantum computer will require an exponential number of gates as the number of qubits $N$ increases, since an element in the Lie algebra $\su[2^N]$ is described by $4^N-1$ parameters in general. 
Nevertheless, we describe an approach for implementing the Riemannian gradient in a circuit in order to set up an approximate scheme that requires only a polynomial number of operations. 


An exact approach to implement the Riemannian gradient on a quantum circuit is to decompose the skew-Hermitian operators $\comm{U_k \rho_0 U_k^\dag}{H}$ in terms of an basis of the Lie algebra $\su[2^N]$. One such basis is the set of Pauli words $\mathcal{P}^{N} = \{P^j\}$ multiplied by $i$ to ensure skew-Hermiticity. 

\begin{figure}[htb!]
    \centering
    \includegraphics[width=\columnwidth]{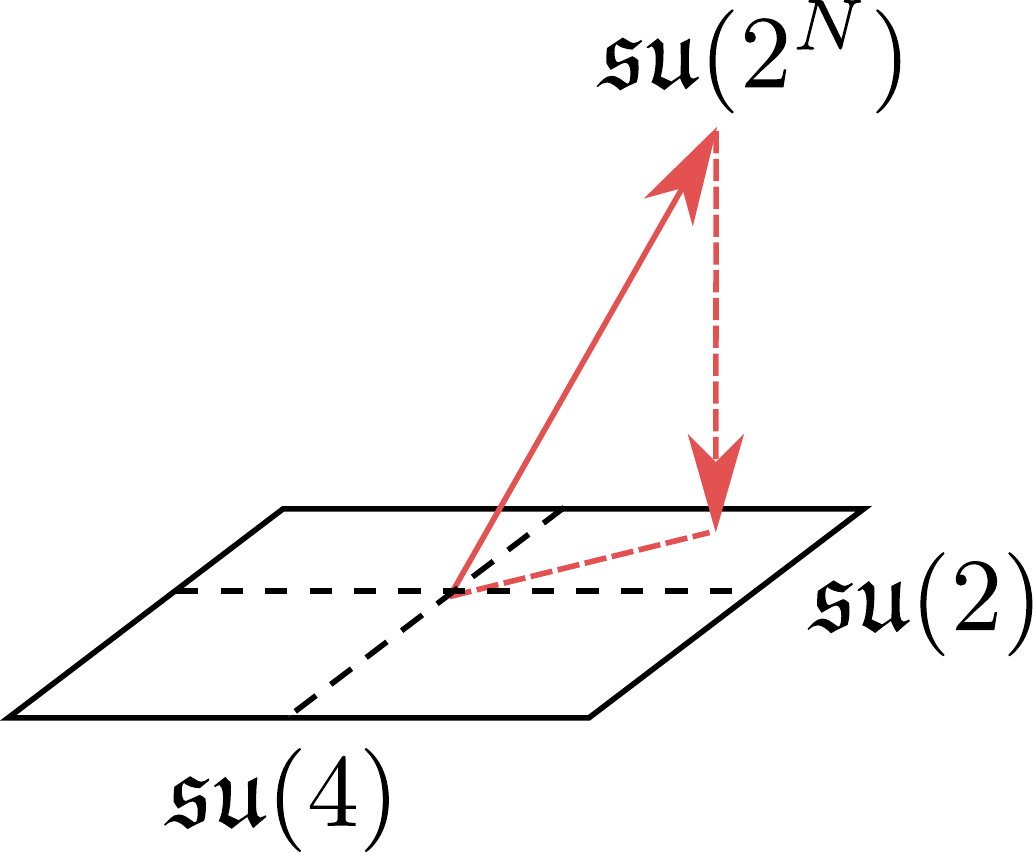}
    \caption{Restricting the algebra to a subspace and projecting the Riemannian gradient onto this subspace. Schematically, one can also break down the projected subspace into further component subspaces (represented for simplicity as single lines).}
    \label{fig:proj}
\end{figure}

We can write the commutator in the exponent of \Cref{eq:discr_riemannian_gradient_flow} in terms of this basis,
\begin{align}
    \comm{U_k \rho_0 U_k^\dag}{H} = -\frac{1}{2^N}\sum_{j=1}^{4^N-1} \Tr{\comm{U_k \rho_0 U_k^\dag}{H} P^j}P^j \label{eq:proj}.
\end{align}
The coefficients 
\begin{align}
    \omega_{k}^j = \Tr{\comm{U_k \rho_0 U_k^\dag}{H} P^j} = \expval{\comm{H}{P^j}}_{\rho_k} \label{eq:omegas},
\end{align}
with $\rho_k = U_k \rho_0 U_k^\dag$ can then be calculated on a quantum device with the parameter-shift rule \cite{Mitarai2018grads, Schuld2019grads, Wierichs2021qdiff,Kyriienko2021qdiff}:
\begin{align}
    \expval{\comm{H}{P^j}} &= i \bigg\langle V^\dag(\frac{\pi}{2})HV(\frac{\pi}{2})\nonumber \\
    &-V^\dag(-\frac{\pi}{2})HV(-\frac{\pi}{2})\bigg\rangle_{\rho_k} \label{eq:omegas_ps},
\end{align}
with $V(t) = \exp\{i t P^j/2\}$ and the expectation value is calculated with respect to the state $U_k \rho_0 U_k^\dag$. Hence estimating the coefficients $\omega_{k}^j$ requires taking the gradient of $\expval{H}_t$ with respect to $t$ given the state $V(t)U_k\ket{\psi_0} $.
The resulting Riemannian gradient flow can be compactly written as
\begin{align}
    U_{k+1} \approx
     \prod_{j=1}^{4^N-1} \exp{-\epsilon \omega_{k}^jP^j}U_k, \label{eq:proj_gradflow}
\end{align}
where absorbed the exponential factor into $\epsilon$ and took the sum out of the product via the Trotter formula at the cost of an error of $\mathcal{O}(\epsilon^2)$. In addition to requiring $4^N-1$ estimates of $\omega_{k}^j$, this also requires applying the corresponding multi-qubit gates generated by all Pauli words of size $N$, which will be very difficult in practice.

Note that instead of splitting the exponent of the sum with a Trotter decomposition, we could directly use a Cartan decomposition algorithm, e.g., the Khaneja-Glaser or D'Alessandro decomposition to recursively decompose the Riemannian gradient into products of single- and two-qubit unitaries \cite{Khaneja2001cartan,Earp2005decomp, Dalessandro2006decompuni}.  

\section{Approximate Riemannian gradient flow in quantum circuits \label{sec:approx}}

To circumvent the exponential resources required for the exact Riemannian gradient, we consider an approximation scheme that requires only a polynomial number of parameters and gates. A natural approximation is restricting the Riemannian gradient to a subspace $\mathfrak{k}\subseteq \su[2^N]$ via an orthogonal projection onto $\mathfrak{k}$. We show this schematically in \Cref{fig:proj}. If we let $\{K^j\}\subset \mathcal{P}^N$ for $j=1,\ldots,k$ be a basis of the subspace $\mathfrak{k}$, then from \Cref{eq:proj_gradflow} we obtain the local Riemannian gradient flow 
\begin{align}
    U_{k+1} &\approx \prod_{j=1}^k\exp{-\epsilon \omega_{k}^j K^j} U_k. \label{eq:practical}
\end{align}
where now $\omega_{k}^j = \expval{\comm{H}{K^j}}_{\rho_k}$.
This approximation gives us control over which directions in the Lie algebra we want to explore. Depending on the choice of $\mathfrak{k}$, we append a sequence of $k$ gates at each optimization step.
For example, we could take the subspace to consist only of 2-local Paulis ($k=9N(N-1)$), nearest-neighbor 2-local Paulis ($k=9N$), or generators of single-qubit rotations ($k=3N$). 

Interestingly, the approximate Lie algebra optimization perspective coincides with the standard VQE approach for particular choices of the Lie algebra subspace. For instance, if we restrict the Riemannian gradient to a $\suloc[2](p)$ subalgebra, then we are performing a variant of the circuit structure learning algorithm called Rotosolve, \cite{Mitarai2019rotosolve, Ostaszewski2021rotosolve, Nakanishi2020rotosolve} where instead of minimizing the expectation value $\expval{H}$ per added gate, we follow the Riemannian gradient with a step $\epsilon$. 
Additionally, we can choose the subspace in such a way that the product in \Cref{eq:practical} becomes a two-qubit layer. If we then append only the unitary with the largest $\omega_{k}^j$, we are performing a popular meta-heuristic first introduced in \cite{Grimsley2019adaptvqe} called Adapt-VQE, except we do not re-optimize the parameters of previous layers at each step. Additionally, the Lyapunov control strategy FALQON \cite{Magann2021falqon} can be understood as a Trotterized time evolution where the stepsize of the drift Hamiltonian is set to the Riemannian gradient.


\begin{figure}[htb!]
    \centering
    \includegraphics[width=\columnwidth]{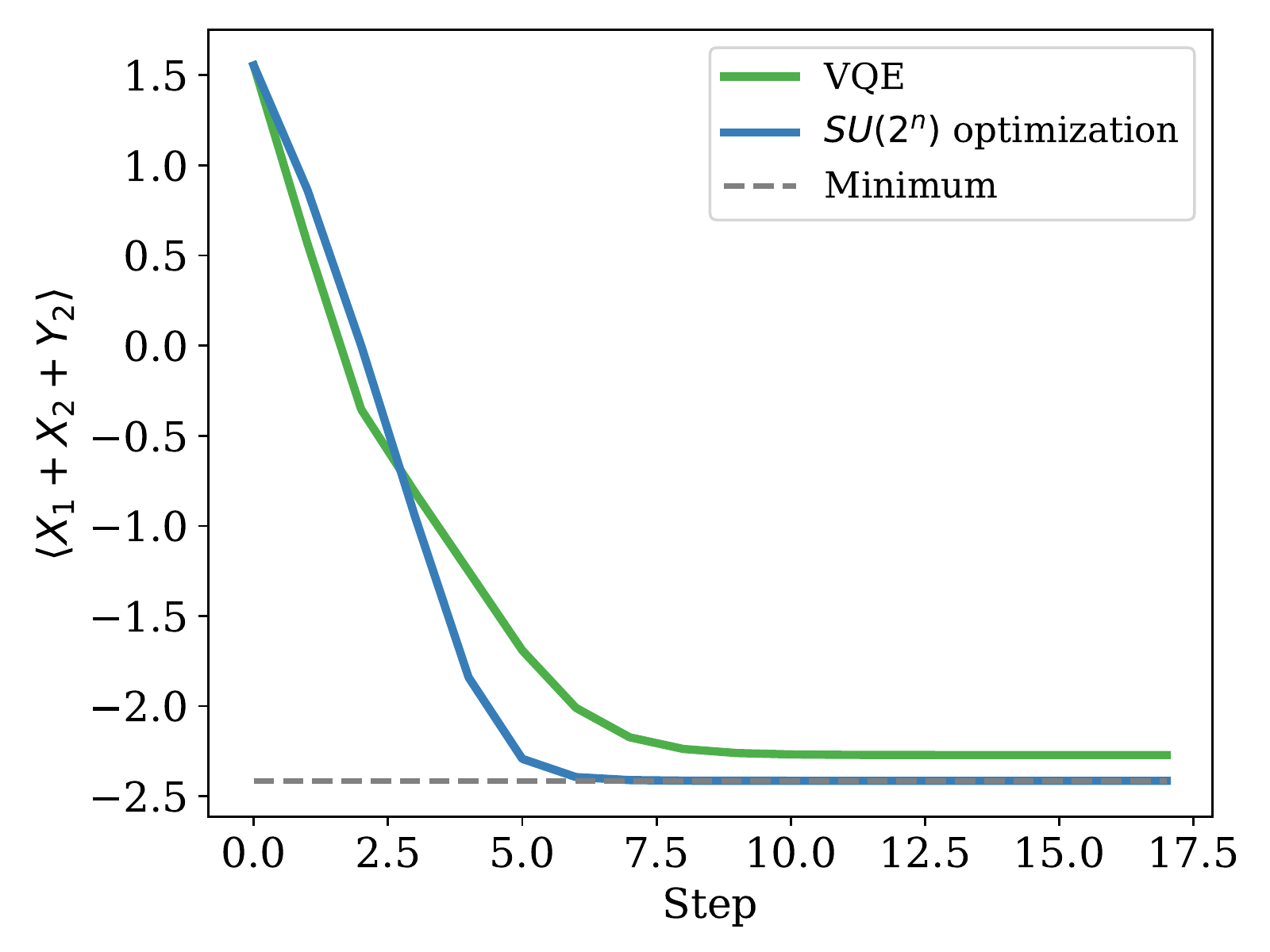}
    \caption{Two-qubit example for the cost function $H = X_1 + X_2 + Y_2$. The circuit at initialization consists of two Hadamards and $\RZ$ gates with parameter $\alpha$ on both qubits, a CNOT (with the control being qubit 1), and finally two $\RZ$ gates with parameter $\beta$ on each qubit. The initial parameters are $(\alpha, \beta) = (0.1,1.2)$. The step size for both the Riemannian gradient and parameter-shift VQE are $\epsilon=0.5$. The VQE optimization gets stuck in a local minimum (also for smaller learning rates), whereas, the Riemannian gradient-flow optimizer rapidly reaches the optimal solution of $\expval{H}\approx-2.40$.}
    \label{fig:diag1}
\end{figure}
With the subspace restriction, the fixed point analysis becomes highly non-trivial. Although we still have the same convergence criterion as before, $\grad \mathcal{L}(U)|_\mathfrak{k}=0$ can be satisfied if the Riemannian gradient only has nonzero components orthogonal to the restricted subspace of the algebra, i.e., $\grad \mathcal{L}(U)|_{\mathfrak{k}} \in \mathfrak{p}$ where $\su[p]=\mathfrak{p}\perp \mathfrak{k}$, and so we lose the global minima guarantees. However, with the right choice of subspace, it is possible that the local Riemannian gradient information is enough to give a good approximation of the global minimum of \Cref{eq:vqe_u}.

Here, we provide several numerical experiments on toy models to test the Riemannian gradient. Our optimization procedure has been implemented in PennyLane as the \texttt{LieAlgebraOptimizer}, which we use for our numerical simulations \cite{pennylane}. First, we consider the exact Riemannian gradient flow, which can be implemented on a circuit for small system sizes. In \Cref{fig:diag1}, we compare the optimizer with the parameter-shift rule for a two-qubit circuit. We see that the Riemannian gradient flow can reach the ground state of a simple Hamiltonian, whereas the VQE optimization can only reach a sub-optimal solution.

\begin{figure}[htb!]
    \centering
    \includegraphics[width=\columnwidth]{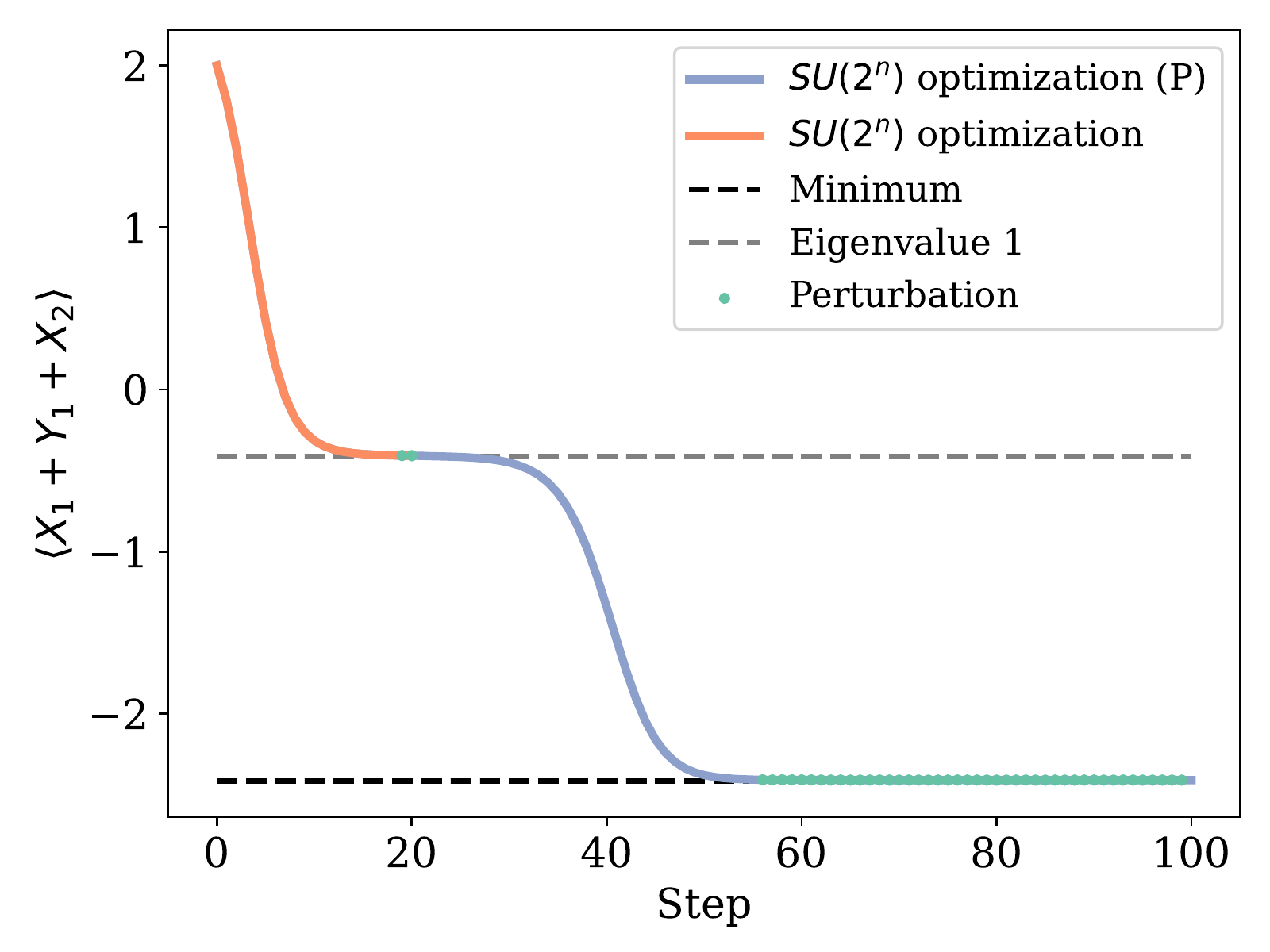}
    \caption{Two-qubit example for the exact Riemannian gradient for the cost function $H = X_1 + Y_1+ X_2$.  The circuit at initialization consists of two Hadamards on each qubit. The learning rate is set at $\epsilon=0.2$. After $20$ steps, the optimization gets stuck in an eigenstate. We generate a stochastic $4\times 4$ matrix $X\sim \mathcal{N}(0, 0.1)^{4\times 4}$ and obtain a random direction in the Lie algebra $K = \frac{i}{2}(X - X^T)$. After 5 perturbations, we escape the saddle point, and the optimization reaches the ground state of $H$.}
    \label{fig:diag2}
\end{figure}

To further illustrate the optimization properties of the Riemannian gradient flow, we study a two-qubit example in \Cref{fig:diag2} where the optimization gets stuck in an eigenstate, which corresponds to a saddle point in the optimization landscape.
After performing a small perturbation in the Lie algebra, we escape the saddle point minimum and converge to the ground state.
   
\begin{figure}[htb!]
    \centering
    \includegraphics[width=\columnwidth]{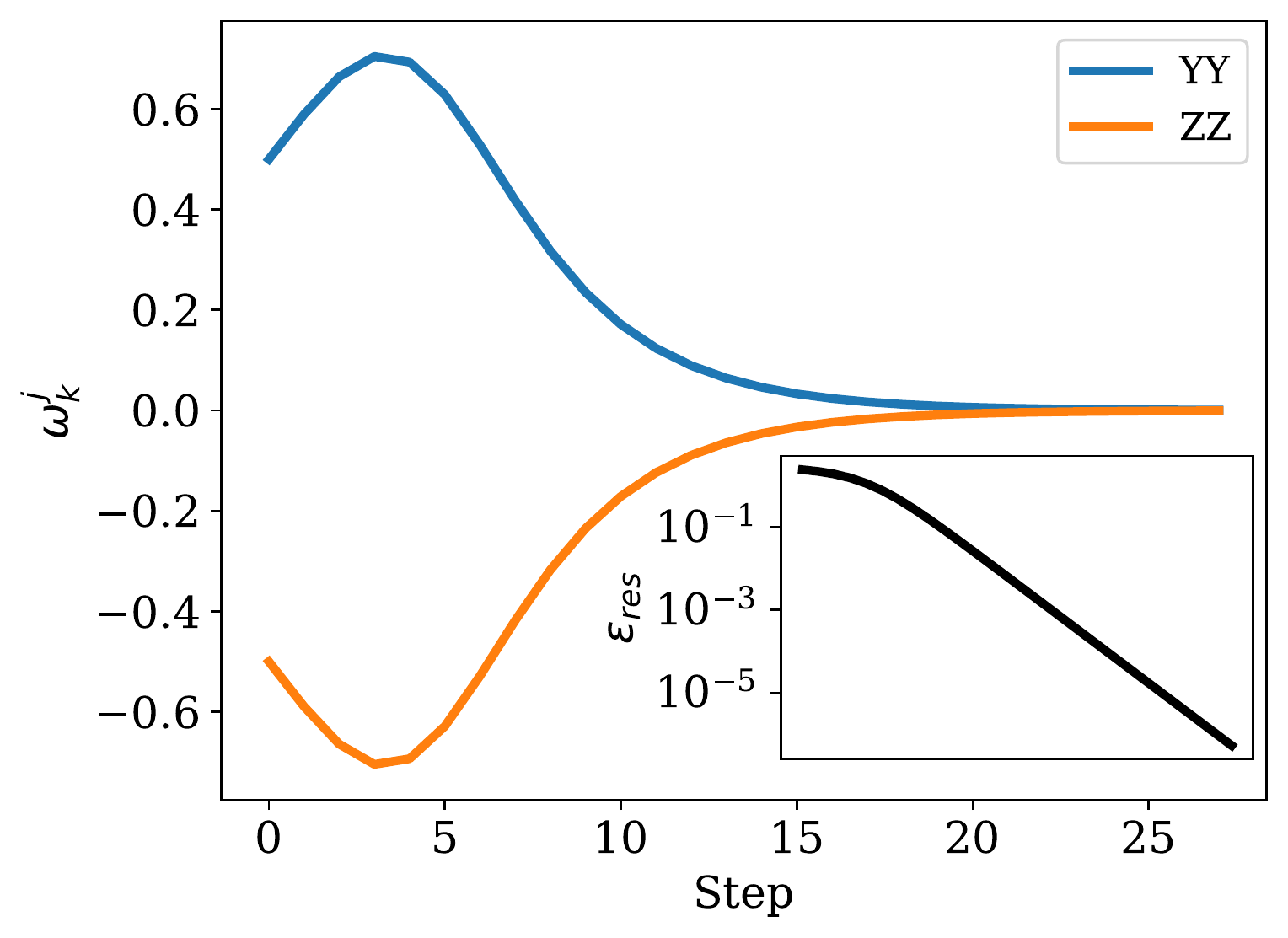}
    \caption{Non-zero components of the Riemannian gradient versus the optimization step. The initial circuit consists of two Hadamard gates. The cost function is $H = X_1 + Y_1 Z_2$. At each step in the optimization, the Riemannian gradient $\grad \mathcal{L}(U) = - \comm{U_k\rho_0 U^\dag_k}{H}$ only has components in the YY and ZZ direction, keeping the state in the submanifold spanned by the states reachable by $(\text{XX}, \text{YY}, \text{ZZ})$. We can therefore restrict the Lie algebra to the subspace $\mathfrak{k}$ spanned by $\{YY,ZZ\}$ and perform the approximate Riemannian gradient flow. At each step, we need to calculate $\{\omega_k^{YY}, \omega_k^{ZZ}\}$. In the inset we see the residual energy $\epsilon_{res} = E_0- \expval{H}$ versus the optimization steps. As the optimization progresses, we get exponentially closer to the ground state of $H$.}
    \label{fig:approx}
\end{figure}

Since the exact Riemannian gradient flow is not experimentally friendly, we consider the approximate optimization scheme from \Cref{sec:approx}. In \Cref{fig:approx}, we see a simple example of the approximate Riemannian gradient flow, where we restricted the full Lie algebra to a subset of directions. We see that after a few steps, we get close to the minimum of the function. For this example, the Lie algebra restriction is still allows us to reach the ground state of the Hamiltonian.

\begin{figure*}[htb!]
    \centering
    \subfloat[\label{fig:e_res}]{
    \includegraphics[width=0.45\textwidth]{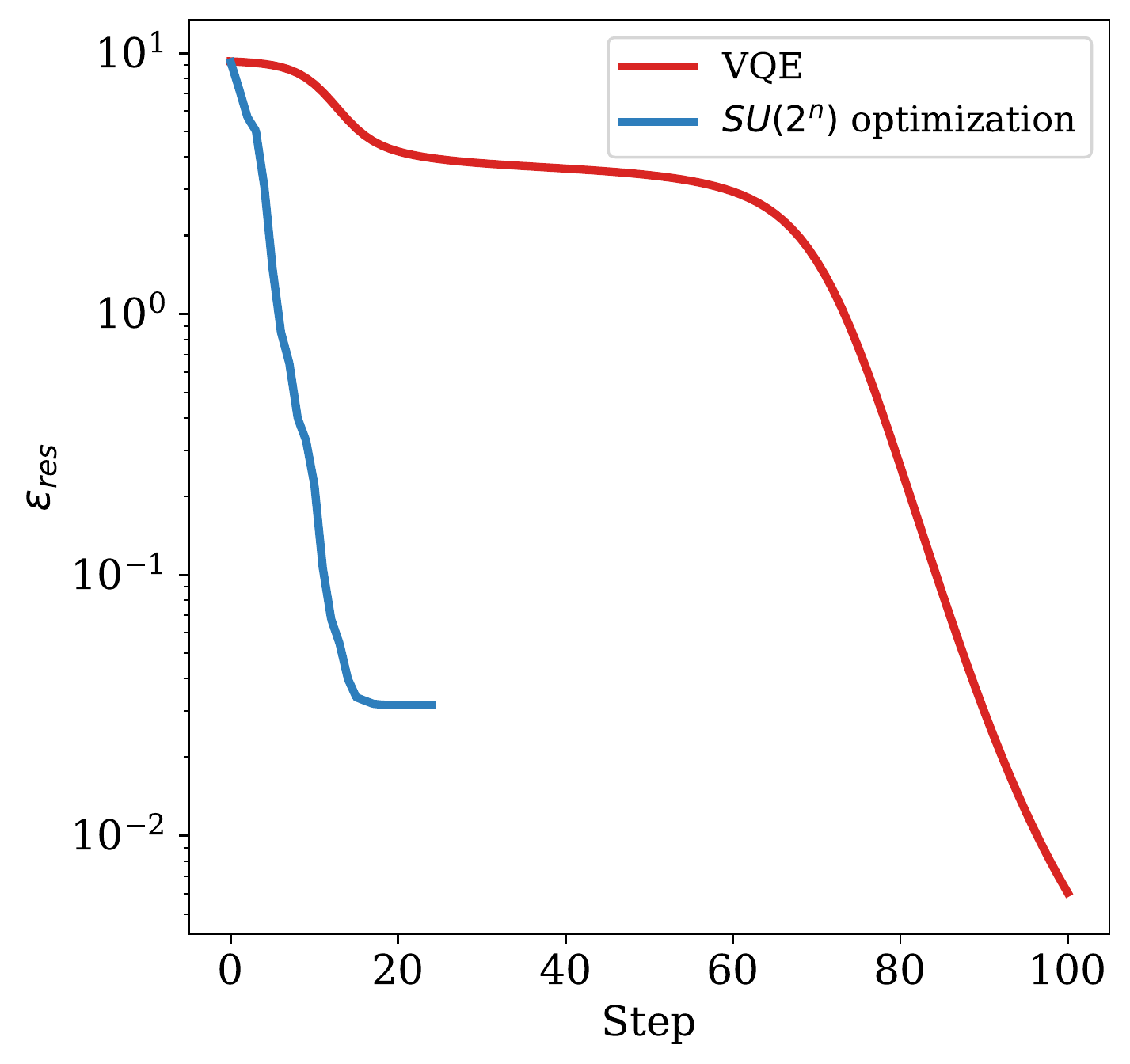}
    }
    \subfloat[\label{fig:directions}]{
\includegraphics[width=0.45\textwidth]{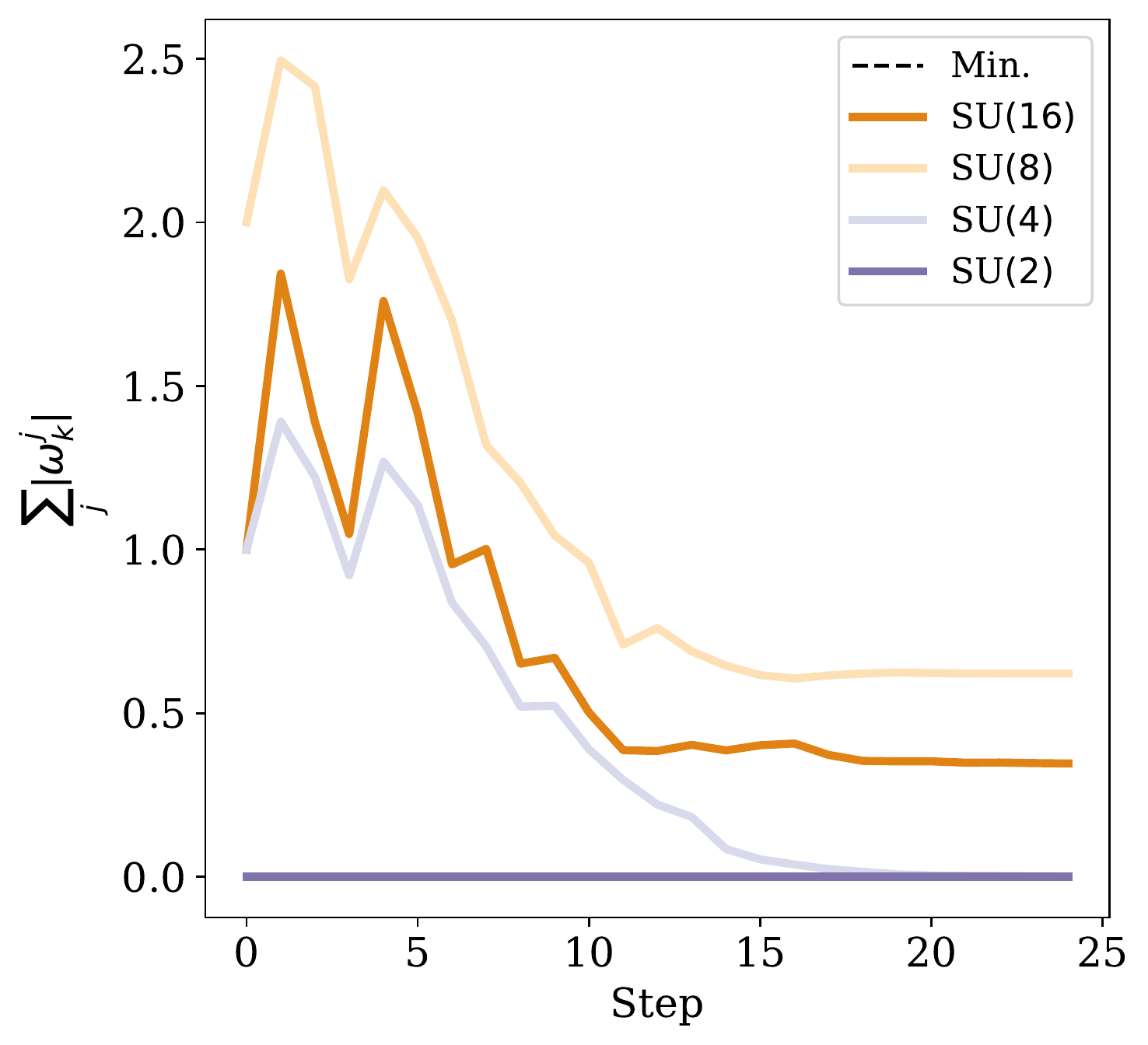}
    }
    \caption{Comparison of Riemannian gradient optimization versus gradient-based VQE for the $4$-qubit transverse field Ising-model. The Riemannian gradient circuit is initialized with a Hadamard on each qubit. To minimize gate costs, we use an adaptive scheme to reduce the amount of gates appended at each step of the Riemannian optimization. We obtain the $\omega_{k}^j$'s on all qubits or pairs of qubits for $\su[2]$ and $\su[4]$, respectively. Then, we select the largest $\omega_{k}^j$ and use we use a structure optimization algorithm to calculate the optimal step size $\epsilon$ \cite{Parrish2019structure, Ostaszewski2021rotosolve, Wierichs2021qdiff, Kyriienko2021qdiff}. The gradient-based VQE optimizer has step size $\epsilon=0.01$. Finally, we append a single gate corresponding to the chosen Lie algebra direction with this step size. (a) The residual energy $\epsilon_{res} = E_0 - \expval{H}$ plateaus for the Riemannian gradient close to the ground state energy. We verify that the optimizer is not stuck in an eigenstate close to the ground state, and so the optimization gets stuck due to the projection of the gradient onto the local algebra. The VQE optimization on the other hand is still getting closer to the ground state. (b) Here, we plot the magnitude of all components of the Riemannian gradient versus the optimization steps. We see that Riemannian gradient becomes zero in the $\su[4]$ direction, but higher order Lie algebra directions are still non-zero. This explains why we cannot converge close to the ground state: we need to access higher order elements of the Lie algebra. }
    \label{fig:tfim}
\end{figure*}

Although the local approximation provides an accurate solution for the previous toy example, we can run into issues for more non-trivial problems, as we see in the final example. We consider the problem of finding the ground state of the Transverse Field Ising model on four qubits, whose Hamiltonian is given by
\begin{align}
    H = -\sum_i\left( Z_i Z_{i+1} + g X_i\right).
\end{align}
We assume periodic boundary conditions and set g=1. The ground state of this model can be reached with a depth $N/2$ ansatz for an $N$-qubit chain using gradient-based VQE \cite{Ho2019hva, Wierichs2020avoiding, Wiersema2020exploring}. We find that the approximate Riemannian gradient optimizer can get close to the ground state. But unlike standard VQE, we cannot approximate the ground state closer than $1\times 10^{-2}$, as can be seen in \Cref{fig:tfim}.

Here, we see a limitation of the approximate Riemannian gradient flow. If we restrict the Lie algebra to $\su[2]$ and $\su[4]$ operators, the Riemannian gradient only has a local view of the cost landscape, and cannot access higher-order Lie algebra directions. On the contrary, VQE can access these directions, since the ansatz is often universal, i.e., made from a product of single and two-qubit unitaries. In principle, the unitary that is implemented by such an ansatz could have a generator $W(\theta)$ such that
\begin{align}
    U_{\text{VQE}}(\theta) = \exp{-i W(\theta)},
\end{align}
that can explore additional $\su[p]$ directions in the Lie algebra for $p=8,16,\ldots$, albeit with a restricted parameterization. 
A bottleneck for gradient-based VQE is that the number of circuit evaluations per optimization step scales linearly in the number of parameters, which is difficult in practice since parallel evaluation of quantum gradients requires multiple quantum devices. Here the approximate Riemannian gradient flow could provide an advantage over standard VQE in that the amount of circuit evaluations is constant independent of circuit depth: we only require $\abs{\mathfrak{k}}$ gradient calculations at each step. However, the Riemannian gradient flow may produce a circuit that is much deeper than the VQE ansatz, since we are appending gates to the circuit at each step $k$.





\section{Conclusion \label{sec:conclusion}}
In this work, we proposed Riemannian gradient flows in the context of variational quantum circuits. We showed that one can perform these types of optimizations on a quantum circuit, with strong convergence guarantees holding for exponentially deep variants of this algorithm. The power and fixed point analysis of local approximations to the Riemannian gradient flow merits further investigation in order to understand the power of this class of algorithms.
 
We hope that this alternative optimization paradigm will allow for understanding the rapidly growing literature in the field of variational quantum algorithms, and provide new insight for variational algorithms in noisy intermediate-scale quantum hardware. Additionally, we believe that the differential geometry and Lie algebra perspective can be a fruitful direction of research to further our understanding of the optimization properties of both old and new variational quantum algorithms \cite{Larocca2021optcon, Kokcu2021cartan}.

In particular, these ideas could be used to investigate overparameterization in VQE \cite{Larocca2021overparam}. Although one can decompose a $\SU[2^N]$ unitary into a product of local parameterized unitaries, there are no guarantees that one is able to find the ground state of a Hamiltonian with a gradient descent-type optimization, even if the circuit is exponentially deep. However, there is evidence that deep quantum circuits can have favorable optimization properties \cite{kiani2020learning, Wiersema2020exploring,Kim2021overparameterization}. Perhaps the global convergence guarantees of double bracket flows flows can be used to understand the convergence properties of deep quantum circuits and provide deeper insight into the power of VQE optimization.

\section{Acknowledgements}
We would like to thank Maria Schuld and Elies Gil-Fuster  for the discussions and we are grateful to Filippo Miatto for providing helpful references. Our thanks also goes to Cunlu Zhou for proofreading the manuscript. Roeland would like to thank David Wakeham and Jack Ceroni from the I.C. for the weekly discussions on differential geometry and Josh Izaac for his help with implementing Riemannian gradient flow in PennyLane. We acknowledge the funding provided by MITACS for this project.
\bibliographystyle{unsrt}
\bibliography{library.bib}

\onecolumngrid
\renewcommand{\appendixtocname}{Supplementary material}

\renewcommand{\thesection}{\Alph{section}}
\renewcommand{\theequation}{\Alph{section}\arabic{equation}}
\counterwithin*{equation}{section}
\setcounter{section}{0}
\pagebreak
\widetext
\begin{center}
\textbf{\large Supplemental Materials}
\end{center}
\setcounter{equation}{0}
\setcounter{figure}{0}
\setcounter{table}{0}
\setcounter{page}{1}
\makeatletter

\section{Differential geometry}
To establish notation, we briefly summarize some of the key concepts in differential geometry needed for our purposes. There exist many excellent references on the topic, see \cite{Baez1995gauge, Nakahara2003gtp} for the physicist-friendly references and \cite{DoCarmo1992riemann, Lee2003introduction} for the more technical expositions on the subject.
\subsection{Manifolds \label{app:manifolds}}
A space $\M$ is called an $n$-dimensional topological manifold if it is locally homeomorphic to $\R^n$. Specifically there must exist a family of open subsets $U_a\subseteq\M$ such that
\begin{enumerate}
    \item the family covers $\M$, i.e., $\bigcup_{a} U_a = \M$,
    \item $\forall a$, $\exists \varphi_a: U_a\to \varphi(U_a) \subset \R^n$ where $\varphi_a$ is homeomorphic.
\end{enumerate}
The pair $(U_a, \varphi_a)$ is called a chart and a collection of charts that covers $\M$ is called an atlas $\mathcal{A}$. In order to develop a differential calculus, we require that all charts in $\mathcal{A}$ are $C^k$-compatible. This means that if we have two charts $(U_a, \varphi_a)$, $(U_b, \varphi_b)$, we require that $\varphi_a(U_a\cap U_b)$ and $\varphi_b(U_a\cap U_b)$ are open and that $\varphi_a \circ \varphi_b^{-1}$ is $C^k$ differentiable. The tuple $(\M, \mathcal{A})$ is called a $k$-differentiable manifold if all charts in $\mathcal{A}$ are $C^k$-compatible. A function $f:\M \to \N$ is said to be $C^k$ differentiable if for all charts $(U_a, \varphi_a)$ on $\M$, $(V_j, \psi_j)$ on $\N$ in the atlas we have that $\psi_j \circ f \circ \varphi_a^{-1}$ is $C^k$ differentiable.

\subsection{Tangent spaces \label{app:tangent_space}}
We are interested in generalizing the concept of a derivative to arbitrary manifolds. Consider a curve  $\gamma: I \to \M$ where $I=(-a,a)$ is an open subset of $\R$ and $\M$ is a differentiable manifold. We can construct a curve on $\M$ so that $\gamma(0)=p$. Then we can ask, what is the derivative of a function $f:\mathcal{M} \to \R$ in the direction of this curve? By working in a chart $(U, \phi)$, $p\in U$ and $\phi(p) = \{x^i\}$ called the coordinate basis where $x^i$ is the $i$th coordinate of the vector $\phi(p)$, we find
\begin{align}
    \frac{df(\gamma(t))}{dt} \bigg|_{t=0} &= \frac{\partial (f\circ\phi^{-1})}{\partial x^i}\frac{d(\phi\circ\gamma)(t)}{dt}\bigg|_{t=0} = \frac{\partial f}{\partial x^i}\frac{d x^i(\gamma(t))}{dt}\bigg|_{t=0}.
\end{align}
This allows us to define a tangent vector at $p$ as
\begin{align}
    v = v^i \frac{\partial}{\partial x^i}, \quad v^i = \frac{d x^i(\gamma(t))}{dt}\bigg|_{t=0}.
\end{align}
So a tangent vector is a an operator that differentiates a function in the direction of some curve $\gamma(t)$ going through a point $p$ as $v(f)(p)$. There exist many such curves, and these curves form an equivalence class. The collection of these equivalence classes is called the tangent space $T_p \M$ of $\M$ at $p$. The tangent space is then a vector space over linear maps called tangent vectors $v: C^\infty(\M) \to \R$, and can be spanned by a basis of differential operators $\{\partial/\partial x_i\}\equiv\{\partial_i\}$. 
Since $T_p \M$ is a vector space, there exists a dual vector space $T^*_p \M$ called the cotangent space. Elements of the cotangent space are called cotangent vectors or one-forms $\omega: T_p \M \to\R$, which accept a tangent vector and produce real number. A one-form can be expanded into a basis that is dual to $\{\partial_i\}$,
\begin{align}
    \omega = \omega_i dx^i,
\end{align}
where $\partial_i dx^j = \delta^i_j$. The most important one-form for our purposes is the (exterior) derivative or differential $df$ that takes a function and creates a one-form. The action of $df$ is defined as
\begin{align}
    d f (v) &= v(f)\\
    df &= \partial_i f(x^1,\ldots,x^n) dx^i
\end{align}

\subsection{Riemmanian Manifolds \label{app:riemannian_manifolds}}
A Riemannian manifold is a manifold $\M$ equipped with a symmetric, non-degenerate metric $g:T_p\M \times T_p \M \to \R$. Given a basis $\{dx^i\}$ on $T_p^* \M$, the metric can be written as
\begin{align}
    g = g_{ij} dx^i \otimes dx^j \label{eq:riemmanian_metric}.
\end{align}
The metric thus defines an inner product between tangent vectors, which we denote by $\expval{,}$.
This inner product induces an isomorphism ${\flat: T_p \M \to T_p^* \M}$ called a musical isomorphism,
\begin{align}
    \flat(v) = \expval{v,.}, \quad \forall v \in T_p \M
\end{align}
with corresponding inverse, ${\sharp: T_p^* \M \to T_p \M}$ given by ${\sharp(v) = \flat(v)^{-1}}$, $\forall v \in T_p \M$. If we choose a basis $\{\partial_i\}$ on $T_p\M$ such that $\partial_i dx^j =\delta_i^j$, we see that 
\begin{align}
    \flat(v) &= \expval{v^k \partial_k, .}\\
    &= v^k (g_{ij} dx^i \otimes dx^j) \partial_k\\
    &= v^i g_{ij} dx^j
\end{align}
and so
\begin{align}
    \sharp(dx^i) = g^{ij} \partial_j,
\end{align}
so that $\sharp(\flat(v)) = v$. Hence the metric allows us to convert tangent vectors into one-forms and vice versa. More importantly, the metric allows us to talk about distance and angles and provides a natural way to generalize the idea of a gradient to a Riemannian manifold. Remember from the previous section that the differential $df$ creates a one-form from a tangent vector. If we define 
\begin{align}
    \sharp(df (v)) \coloneqq \grad f \label{eq:def_riemann_grad}
\end{align}
to be the Riemannian gradient with respect to the metric $g$, then the differential of a vector $v\in T_p\mathcal{M}$ can be written as
\begin{align}
    df(v) = \expval{\grad f, v}.
\end{align}
We can see that, by construction, the Riemannian gradient is an element of the tangent space, $\grad f \in T_p \mathcal{M}$, since the $\sharp$ operation produces a tangent vector. Additionally, $\grad f$ is perpendicular to the level curves at each point $x\in\mathcal{M}$ under the metric. To see this, consider a tangent vector $v$ that points along the level curves of $f$, clearly we then have $df(v)=0$ and thus $\grad f \perp v$. 

Note that if we take the metric $g_{ij}$ to be the standard Euclidean metric in the standard coordinate basis, we recover the gradient from multi-variable calculus:
\begin{align}
    \sharp(df (v)) = (\partial_i f) v^i = \nabla f \cdot v.
\end{align}
Since $df(v)$ is metric-independent, we can understand the construction of the Riemannian gradient as requiring that $df(v) = {\expval{\grad f, v} \equiv \nabla f \cdot v}$ in the standard chart. This is called the compatibility condition of the Riemannian gradient.

\section{\texorpdfstring{The group $\SU[p]$\label{app:su_p}}{The group SU[p]}}
Consider the special unitary Lie group $\SU[p]$:
\begin{align}
    \SU[p] \coloneqq \{ X \in \mathbb{C}^{p\times p} | X^\dag X = I, \det{X}=1\}.
\end{align}
This group is equal to $\text{U}(p)$ up to a global phase, and has dimension $p^2-1$. Consider now a curve $X(t):\R \to \SU[p]$, where $\forall t$, $X^\dag X = I$ and $\det{X}=1$. If we differentiate this condition with respect to $t$, we obtain
\begin{align}
    \frac{d}{dt} (X^\dag(t)X(t)) &= 0,\\
    \dot{X}^\dag(t)X(t) + X^\dag(t)\dot{X}(t) &= 0.
\end{align}
If $X(t)$ passes through $X$ at time $t=0$, then we see that $\dot{X}(0) = V$ must satisfy
\begin{align}
    T_X \SU[p] \coloneqq\{V\in\mathbb{C}^{p\times p} | V^\dag X + X^\dag V = 0\},
\end{align}
so $V$ must be a skew-Hermitian matrix \cite{Fiori2005geodesiclearning}. The Lie algebra is the tangent space of a Lie group at the identity. Hence for $\SU[p]$, 
\begin{align}
    \su[p] \coloneqq \{\Omega\in\mathbb{C}^{p\times p} | \Omega^\dag=-\Omega\}.
\end{align}
We see that the elements $\Omega\in \su[p]$ are related to Hermitian matrices $H$ by $\Omega = iH$. By multiplying elements $X$ of $\SU[p]$ to the right or left with an element of the algebra, we can move from the tangent space at the identity to the tangent space at $X$
\begin{align}
    T_X \SU[p] \coloneqq \{V = \Omega X| \Omega \in \su[p]\}.
\end{align}

\section{Riemannian gradient flow }
The following is due to \cite{SchulteHerbruggen2010gradflow}. 
\subsection{\texorpdfstring{$\SU[p]$ flow}{SU[p] gradient flow} \label{app:lie_grad_flow}}
For $\SU[p]$ there exists a bi-invariant metric $\expval{,}: T_X \SU[p]\times T_X \SU[p] \to \R$ that induces a Riemannian gradient on the group \cite{Arvanitoyeorgos2003AnIT}. This bi-invariant metric is given by $\expval{WX,VX} = \Tr{X^\dag W^\dag V X} = \expval{W,V}$, $\forall W,V\in T_X \SU[p]$. Consider the function
\begin{align}
    h :\SU[p] \to \mathbb{C}^{p\times p}, \quad h(X) \coloneqq C^\dag X A X^\dag, \label{eq:h_su(p)}
\end{align}
where $C$ and $A$ are Hermitian matrices on $\mathbb{C}^{p \times p}$. For a tangent vector $\Omega X \in T_X \SU[p]$, the derivative of $h$ is 
\begin{align}
    d\: h(X)(\Omega X) &= \left(C^\dag (dX) A X^\dag + C^\dag X A (dX)^\dag\right) (\Omega X)\\
    & =  C^\dag \Omega X A X^\dag - C^\dag X A X^\dag \Omega.
\end{align}
Because $dX(\Omega X) = \Omega X $. The derivative of $f:\SU[p]\to \R$,  $f(X) = \Tr{h(X)}$ at $\Omega X$ is
\begin{align}
    d\: f(X) (\Omega X) = \Tr{d\:h(X) (\Omega X)},
\end{align}
by the linearity of the trace. Defining $\tilde{A} \coloneqq X A X^\dag$ we find
\begin{align}
d\: f(X) (\Omega X)      & = \Tr{C^\dag \Omega \tilde{A} - C^\dag \tilde{A} \Omega}\\
    &= \Tr{\comm{\tilde{A}}{C^\dag} \Omega}\\
    & = \expval{\comm{\tilde{A}}{C^\dag}^\dag, \Omega}\\
    & = \expval{-\comm{\tilde{A}}{C^\dag}X, \Omega X},
\end{align}
where we used that $\expval{V,W}=\expval{V X,W X}$ in the final line. We can now identify the Riemannian gradient from the compatibility condition,
\begin{align}
    d\: f(X) (\Omega X) & = \expval{\grad f(X), \Omega X},
\end{align}
so that
\begin{align}
    \grad f(X) = -\comm{\tilde{A}}{C^\dag} X.
\end{align}
Plugging in $\tilde{A} = U \rho_0 U^\dag$ and $C = H$ and flipping the sign to find the minimum of \Cref{eq:vqe_u} instead of the maximum gives the Riemannian gradient flow of \Cref{eq:riemannian_gradient_flow}.

\end{document}